\documentclass{aastex63}

\usepackage{bm}
\submitjournal{ApJL}

\shorttitle{MHD waves in inner heliosphere}
\shortauthors{Zhu et al.}

\graphicspath{{./}{figures/}}

\begin{document}
\title{Wave Composition, Propagation, and Polarization of MHD Turbulence within 0.3AU as Observed by PSP}

\correspondingauthor{Jiansen He}
\email{jshept@pku.edu.cn}

\author[0000-0002-1541-6397]{Xingyu Zhu}
\affiliation{School of Earth and Space Sciences, Peking University \\
No.5 Yiheyuan Road, Haidian District \\
Beijing, 100871, China}

\author[0000-0001-8179-417X]{Jiansen He}
\affiliation{School of Earth and Space Sciences, Peking University \\
No.5 Yiheyuan Road, Haidian District \\
Beijing, 100871, China}

\author[0000-0002-0497-1096]{Daniel Verscharen}
\affiliation{Mullard Space Science Laboratory, University College London, Dorking RH5 6NT, UK}
\affiliation{Space Science Center, University of New Hampshire, Durham NH 03824, USA}

\author[0000-0002-6300-6800]{Die Duan}
\affiliation{School of Earth and Space Sciences, Peking University \\
No.5 Yiheyuan Road, Haidian District \\
Beijing, 100871, China}

\author[0000-0002-1989-3596]{Stuart D. Bale}
\affil{Physics Department, University of California, Berkeley, CA 94720-7300, USA}
\affil{Space Sciences Laboratory, University of California, Berkeley, CA 94720-7450, USA}
\affil{The Blackett Laboratory, Imperial College London, London, SW7 2AZ, UK}
\affil{School of Physics and Astronomy, Queen Mary University of London, London E1 4NS, UK}

\begin{abstract}
Turbulence, a ubiquitous phenomenon in interplanetary space, is crucial for the energy conversion of space plasma at multiple scales. This work focuses on the propagation, polarization and wave composition properties of the solar wind turbulence within 0.3AU, and its variation with heliocentric distances at MHD scales (from 10s to 1000s in the spacecraft frame). We present the probability density function of propagation wavevectors (${\rm{PDF}}(k_\parallel,k_\perp)$) for solar wind turbulence winthin 0.3 AU for the first time: (1) wavevectors cluster quasi-(anti-)parallel to the local background magnetic field for $kd_{\rm i}<0.02$, where $d_{\rm i}$ is the ion inertial length; (2) wavevectors shift to quasi-perpendicular directions for $kd_{\rm i}>0.02$. Based on our wave composition diagnosis, we find that: the outward/anti-sunward Alfv\'en mode dominates over the whole range of scales and distances, the spectral energy density fraction of the inward/sunward fast mode decreases with distance, and the fractional energy densities of the inward and outward slow mode increase with distance. The outward fast mode and inward Alfv\'en mode represent minority populations throughout the explored range of distances and scales. On average, the degree of anisotropy of the magnetic fluctuations defined with respect to the minimum variation direction decreases with increasing scale, with no trend in distance at all scales. Our results provide comprehensive insight into the scenario of transport and transfer of the solar wind fluctuations/turbulence in the inner heliosphere.

\end{abstract}
\keywords{}

\section{Introduction} \label{sec:intro}
The solar corona dynamically expands into interplanetary space in the form of the continuous solar wind \citep{Parker1958,Coleman1966}, the birth and the acceleration mechanism of which are still not well understood \citep{Tu2005,He2007,Cranmer2019}. The solar wind flowing into interplanetary space carries information about its source region, and on the other hand, involves diversified nonlinear physical processes \citep{Tu1995,Bruno2013}. It is essential to investigate the nature of near-sun fluctuations in order to analyze and understand these nonlinear physical processes as well as the heating and acceleration mechanisms of solar wind.

The statistical properties of the solar wind generally vary with speed, location and type of source region and heliocentric distance \citep{Bavassano1982,Tu1989,He2013,Matteini2014,Stansby2018,Horbury2018,Wang2019,Perrone2019,Bandyopadhyay2020,Chen2020,Chhiber2020,Duan2020,Qudsi2020}. \citet{Tu1989} contrast the properties of MHD turbulence between high speed and low speed solar wind at 0.3AU using the spectra of Els\"asser variables, cross helicity, residual energy, Alfv\'en ratio and Els\"asser ratio. They consider that, compared to the high speed wind, the turbulence evolves in an advanced state in slow wind, due to the longer expansion time. The $z^+$ and $z^-$ are close to a balanced state with an approximate -1.67 spectral index. The mode composition therein is dominated by the Alfv\'en mode and slow mode in the limit of incompressibility \citep{Dobrowolny1980}. \citet{Bavassano1982} study the variation of the nature of the fluctuations with heliocentric distance and scale in the trailing edge of a stream interaction region. The anisotropy defined with respect to the direction corresponding to the minimum eigenvalue decreases as the heliocentric distance increases and the scale decreases. The magnetic field closer to the sun is more compressed. However, \citet{Chen2020} report the evolution of solar wind turbulence from 0.17AU to 1AU, recently, and find at 0.17AU that: (1) the spectra of magnetic field, velocity and Els\"asser variables present a -3/2 slope at MHD scales; (2) the magnetic field is less compressed and (3) the outward propagating Alfv\'en waves are more dominant than at 1AU. Fast solar wind is characterised by highly Alfv\'enic fluctuations, although a new type of Alfv\'enic slow solar wind, possibly coming from quiet-Sun regions or coronal-hole boundaries, has been reported at distances from 0.3 AU to 1 AU. \citep{D'Amicis2015,Wang2019,Perrone2020,Parashar2020}. It is of interest to study this kind of solar wind, on account of its distinct properties, which differ from the classical slow solar wind. 

Even if we relax the assumption of a pure superposition of linear waves, nonlinear turbulent fluctuations still retain certain polarization and correlation properties of linear modes \citep{Tu1995}. When we use the term 'wave', we refer to the mode composition of the fluctuations within this wave-turbulence paradigm. The composition of wave modes in the solar wind near 1 AU has been extensively studied and controversially discussed. There are many means to diagnose the wave modes: correlation analysis between velocity and magnetic field fluctuations \citep{Wang2012,Safrankova2019}, cross helicity analysis \citep{Roberts1987}, comparison of the MHD dispersion relations derived from measurements with theory predictions \citep{Shi2015}, and mode recognition methods \citep{Glassmeier1995,Narita2015,Chaston2020}. According to these studies, non-compressive outward Alfv\'en modes dominate the fluctuations in the solar wind especially in fast streams \citep{Bruno2013}. Compressive waves likely suffer strong Landau damping \citep{Barnes1966}, resulting in their suppression in the overall fluctuations. Correlations among variables (e.g. magnetic pressure, thermal pressure, density and temperature), show that the compressive component simultaneously exists of magnetosonic waves and pressure balanced structures (PBSs) \citep{Kellogg2005,Yao2011,Yang2017}. The majority of the compressive fluctuations is slow-mode-like rather than fast-mode-like in polarization \citep{Howes2012,He2015,Shi2015}. In situ observations show that the Alfv\'enicity decreases with heliocentric distance, which might be caused by the increased contribution of inward propagating Alfv\'en waves or the compressive fluctuations \citep{Bruno1993}.

To further comprehend the underlying multi-scale nature and evolution of near-sun turbulence, we systematically study the variation of the fluctuations' properties with scale and heliocentric distance within 0.3AU. The properties include the propagation direction of the wave, the mode composition, and the characteristic of anisotropy on average. In Section \ref{sec:data}, we briefly introduce the data sets we use. In Section \ref{sec:result}, we present our methods and analysis results, and give our summary and conclusions in Section \ref{sec:sum}. Our observations can provide observational evidence for the verification of existing theoretical models at closer heliocentric distances, and also impose constraints on the improvement of existing theoretical models and the proposal of new models.

\section{ Data Sets and Data Deduction} \label{sec:data}
We conduct our analysis using the data obtained from {\em Parker Solar Probe} (PSP), which is the closest human-built satellite to the sun up to now \citep{Fox2016}. We use the Level-2 magnetic field data supplied by the Flux-gate Magnetometer (MAG; \cite{Bale2016}) and the Level-3i proton data provided by the Solar Probe Cup (SPC; \cite{Kasper2016}). The time interval investigated spans from UTC2018-10-31/20:00:00 to UTC2018-11-10/15:00:00 in which period PSP cruised between 0.166AU (35.78 solar radii)and 0.243AU (56.37 solar radii). The interval we choose is shorter than the high-cadence interval around the first perihelion, because there are several sampling gaps longer than 30 minutes in the other intervals from which the SPC data are unavailable. We analyze time periods of the fluctuations in the range from 10s to 1000s, corresponding to MHD scales in the plasma frame. We do not exclude the so-called `switchback' patterns that exist among various scales (see \cite{Bale2019,Kasper2019,Dudokdewit2020}).

For the analysis of propagation direction and fluctuation anisotropy, we use the Singular Value Decomposition (SVD) method to resolve the frequencies and wavevectors of the waves based on Faraday's law. We estimate the three singular values of the spectral matrix (Eq.(8) of \citet{Santolik2003}), based on the principle of divergence-free magnetic field. We estimate the electric field at MHD scales for our SVD analysis as $E=-V_i\times B$ \citep{Shi2015}, where $V_i$ is the proton bulk velocity obtained from SWEAP/SPC. Note that only one wavevector is solved for every specific frequency with the SVD method. Therefore, the resolved frequency and wavevector can be regarded as the frequency and wavevector of the major wave mode. In reality, it is possible that multiple wave modes exist in the turbulence at the same time and scale. For the mode composition analysis, we use the method suggested by \citet{Glassmeier1995}, and get the contributions of the six MHD modes (parallel and anti-parallel propagating Alfv\'en mode, fast mode and slow mode) to the fluctuations. We estimate the spectral energy density of each mode as $e_i^TS(f_{sc},t)e_i$, where $S(f_{sc},t)$ is the spectral density matrix as defined by \citet{Glassmeier1995} and $e_i$ is the eigenvector of the corresponding mode.

\begin{figure}
\centerline{\includegraphics[width=20cm, clip=]{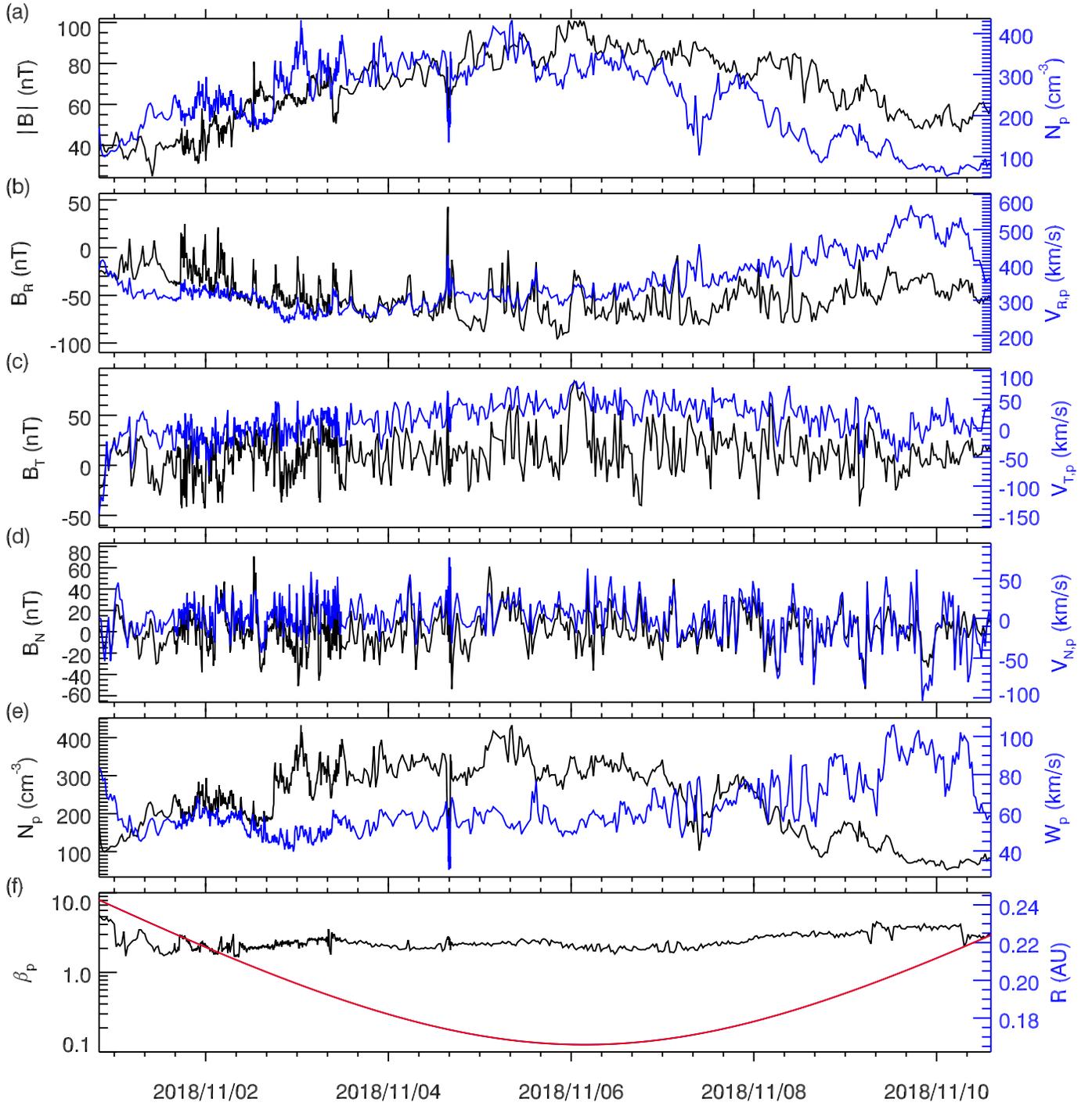}}
\caption{Time sequences overview of magnetic and plasma measurements during PSP's first encounter. Panel (a): magnetic field strength ($|B|$) and proton density ($N_{\rm p}$). Panel (b\&c\&d): magnetic fields ($B_{\rm R}, B_{\rm T}, B_{\rm N}$) and proton bulk velocities ($V_{\rm R,p}, V_{\rm T,p}, V_{\rm N,p}$) in RTN coordinates. Panel (e): proton density ($N_{\rm p}$) and thermal velocity ($W_{\rm p}$). Panel (f): plasma beta ($\beta$) and heliocentric distance ($R$) of spacecraft's position.
\label{fig:fig1}}
\end{figure}

\section{Analysis Results} \label{sec:result}
We present an overview of the observed magnetic field and plasma measurements in Figure \ref{fig:fig1}. To highlight the correlated fluctuations of the variables over such a long duration of about 10 days, we smooth all the measurements with a running window of 30 min. Figure 1(a) shows that the proton density ($N_{\rm p}$) and the magnetic field strength ($|B|$) decrease with increasing heliocentric distance. The three components of the magnetic field ($B_{\rm R}, B_{\rm T}, B_{\rm N}$) and the proton velocity ($V_{\rm R,p}, V_{\rm T,p}, V_{\rm N,p}$) in the RTN coordinates are positively correlated, respectively, which suggests that the large-scale outward-propagating Alfv\'enic fluctuations are dominant during this encounter. The proton thermal velocity ($W_{\rm p}$) varies between 50 km/s and 100 km/s and there is no global correlation between the proton density and the thermal velocity. The plasma beta ($\beta_{\rm p}$) is around 2, which does not exhibit a significant variation with heliocentric distance ($R$).

\begin{figure}
\centerline{\includegraphics[width=20cm, clip=]{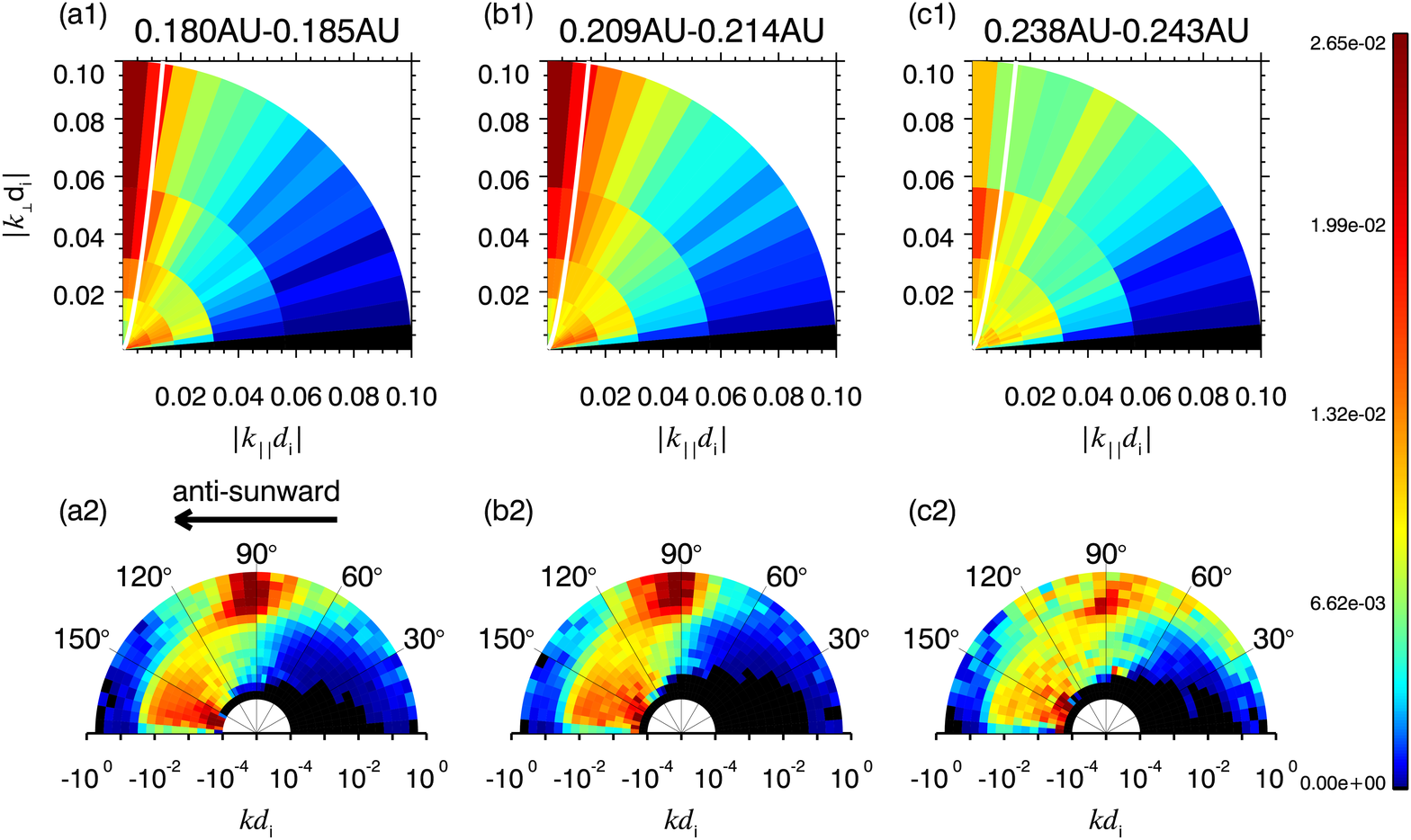}}
\caption{Panel (a1\&b1\&c1): Probability distribution functions of the wavevector in $|k_{\parallel}d_{\rm i}|-|k_{\perp}d_{\rm i}|$ space in the range of 0.180AU-0.185AU, 0.209AU-0.214AU, and 0.238AU-0.243AU, respectively. The white solid lines represent the relation between $k_{\parallel}$ and $k_{\perp}$ as predicted from the phenomenology of critical balance, $k_{\parallel}\sim k_{\perp}^{\frac{2}{3}}k_0^{\frac{1}{3}}$, where $k_0$ is the wavenumber of the outer scale. Panel(a2\&b2\&c2): PDFs of the propagation angle ($\theta_{{\rm k,B_0}}$) at differing scales ($kd_{\rm i}$), in the corresponding distance ranges. 
\label{fig:fig2}}
\end{figure}

We solve the wavevector, $\bm{k}(\tau, R(t))$, at different heliocentric distances ($R(t)$) and periods ($\tau$), where $R$ is a function of time ($t$). The local background magnetic field, $\bm{B_0}(\tau, R(t))$, is acquired by Gaussian-weighting of the magnetic-field time series at time $t$, where the width of the Gaussian profile is defined by the period $\tau$ \citep{Podesta2009}. We then calculate the angles between $\bm{k}$ and $\bm{B_0}$, $\theta_{{\rm k,B_0}}(\tau, R(t))$. Figure \ref{fig:fig2} (a1), (b1) and (c1) show the probability distribution functions (PDFs) of wavevectors in three distance ranges. For $kd_{\rm i}>0.02$, the wavevectors cluster around the quasi-perpendicular direction. For $kd_{\rm i}<0.02$, the most probable wavevectors are quasi-parallel, relative to the local background magnetic field. Figure \ref{fig:fig2} (a2), (b2) and (c2) show the PDFs of $\theta_{\rm k,B_0}$ depending on $kd_{\rm i}$, in three different distance ranges. The propagation angles are close to $160^{\circ}$ for $kd_{\rm i}<0.02$, and close to $90^{\circ}$ for $kd_{\rm i}<0.02$. This indicates that the propagation angles are scale-dependent and turn from quasi-parallel at large scales to quasi-perpendicular at small scales.

\begin{figure}
\plotone{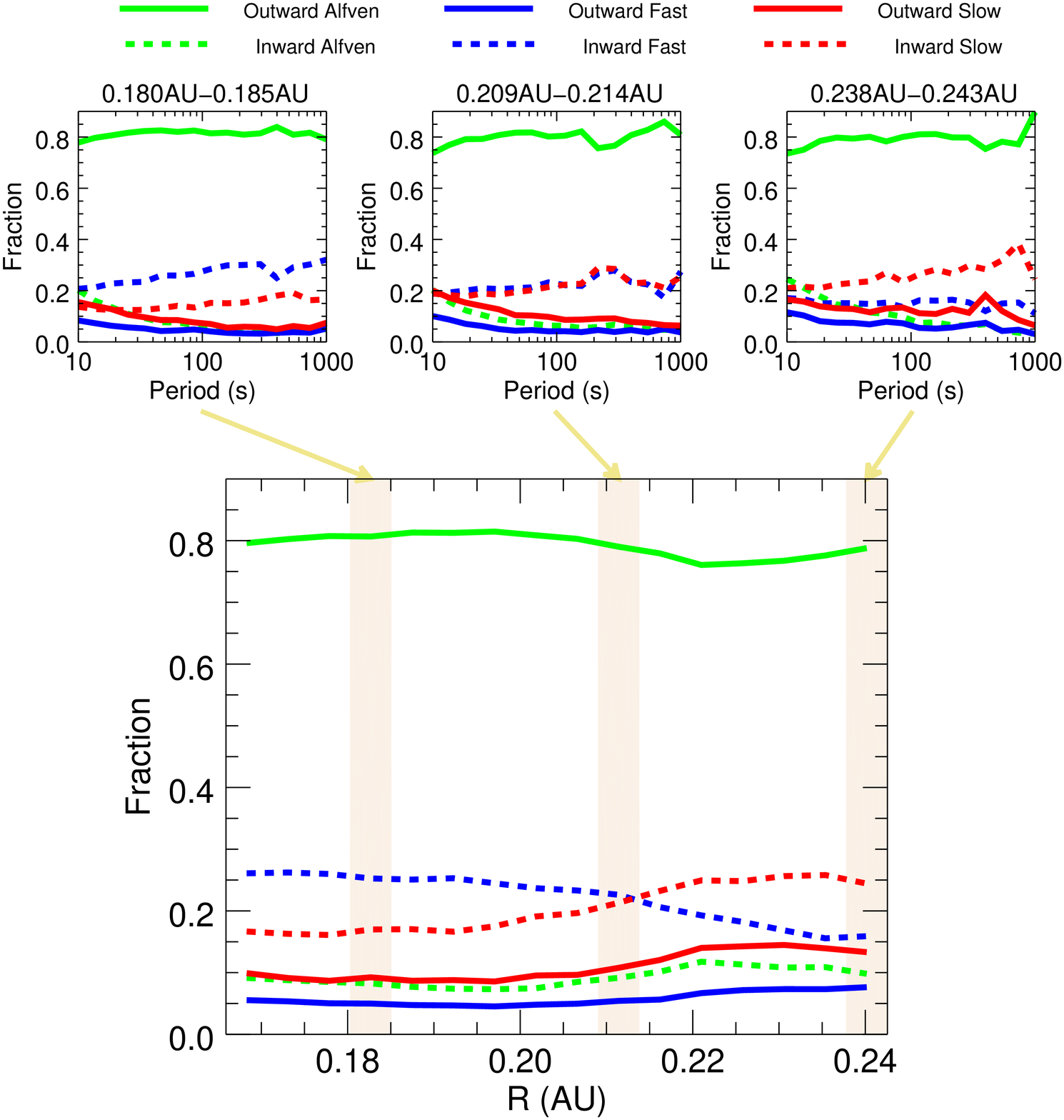}
\caption{(Top) The period-depending variation of the spectral fractions of the six MHD modes i.e., outward/anti-sunward ({\it solid line}) and inward/sunward ({\it dashed line}) propagating Alfv\'en modes ({\it green}), fast modes ({\it blue}) and slow modes ({\it red}), as averaged over different distance ranges: 0.180AU-0.185AU ({\it Left}), 0.209AU-0.214AU ({\it Middle}) and 0.238AU-0.243AU ({\it Right}), respectively. (Bottom) The heliocentric distance variation of the spectral fractions of the six MHD modes as averaged over the time scale (period) from 10 to 1000s. The lime shadow sections correspond to the distance ranges used for the averaging of the intervals in the upper three panels.
\label{fig:fig3}}
\end{figure}

We carry out a mode composition diagnosis \citep{Glassmeier1995} and directly obtain the fractions of the six MHD wave modes, at different heliocentric distances and periods. According to the radial component of the local background magnetic field, $B_{0r}=\bm{B_0}\cdot\bm{\hat{e}_r}$, we transform the parallel and anti-parallel modes into outward/anti-sunward modes when $\bm{k}\cdot\bm{B_0}>0$, and inward/sunward modes when $\bm{k}\cdot\bm{B_0}<0$, respectively. The variation results of the averaged fractions of the transformed MHD modes are shown in Figure \ref{fig:fig3}. The upper three panels show the variation of the fractions of the MHD modes with period averaged over the distance ranges of 0.180AU-0.185AU, 0.209AU-0.214AU and 0.238AU-0.243AU, respectively. The wave mode occupying the highest spectral proportion is the outward Alfv\'en mode at most scales in these three $R$-intervals. The outward fast mode, the inward Alfv\'en mode and the outward slow mode represent the modes with the lowest fractional proportions throughout the whole MHD range at these distances. The fractional proportions of these three modes slightly increase with increasing distance. On average, from 0.180AU to 0.185AU, the inward fast mode is the second-most abundant mode, while the inward slow mode is in third place. From 0.209AU to 0.214AU, the inward slow mode and the inward fast mode have approximately equivalent proportions. From 0.238AU to 0.243AU, the inward slow mode is in second place, followed by the inward fast mode. We also find this change of mode composition with distance in the radial variation of the period-averaged fraction of mode compositions (see lower panel of Figure \ref{fig:fig3}). The outward Alfv\'en mode dominates throughout the whole near-Sun region under investigation. The fractional contribution of the fast mode decreases with increasing distance, while the contribution from inward slow modes increase with distance.

\begin{figure}
\plotone{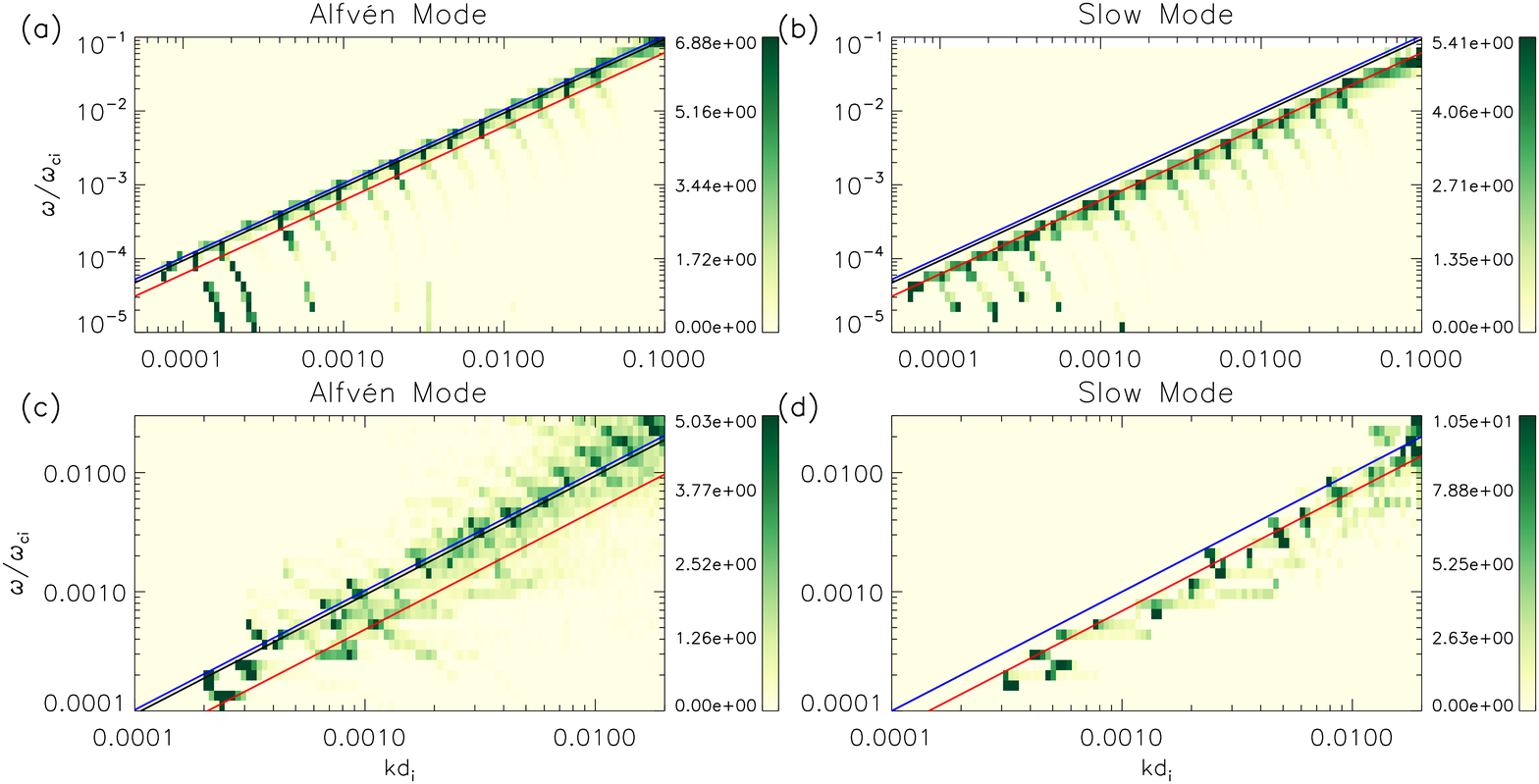}
\caption{Panel (a)\&(b): The PDFs of the normalized wavenumbers, $kd_{\rm i}$, for Alfv\'en waves and slow waves, at each normalized angular frequency, $\omega/\omega_{{\rm ci}}$ in the plasma frame, resolved by a benchmark test of the SVD method, with MHD Alfv\'en-mode and MHD slow-mode fluctuations. The theoretical MHD dispersion relations of Alfv\`en mode, fast mode and slow mode are plotted in black, blue and red solid lines, respectively. Panel(c)\&(d): The PDFs of $kd_{\rm i}$ for Alfv\'en waves and slow waves, at each $\omega/\omega_{{\rm ci}}$, obtained from application of the SVD method to the magnetic and velocity measurements from PSP in [20:00, 21:00] UT on 2018-11-05 (panel c) and [18:20, 18:25] UT on 2018-11-04 (panel d), consistent with the dispersion relations of Alfv\'en and slow modes, respectively. Unlike in panels (a) and (b), we use the averaged plasma parameters over the corresponding time intervals in panels (c) and (d).
\label{fig:fig4}}
\end{figure}

To further verify this composition diagnosis results, we reconstruct the dispersion relations of Alfv\'en waves and slow waves, as shown in Figure \ref{fig:fig4}. We first demonstrate a benchmark test to verify the ability of the SVD method to resolve the MHD dispersion relations. The preset basic parameters are: bulk velocity, $V_0=400{\rm km/s}$, background magnetic field, $B_0=90{\rm nT}$, proton number density, $n_p=300{\rm cm^{-3}}$, proton thermal velocity, $60{\rm km/s}$, $\theta_{{\rm k,B_0}}=20^{\circ}$. Based on the polarization relations of the Alfv\'en mode and slow mode, we set up the corresponding magnetic field and velocity disturbances of the two modes respectively, and use these disturbances as an artificial data input of the SVD method. In order to test the robustness of the SVD method, we also add 0.1\% level of noise for each wave at all scales to the virtual data input. As a result, we obtain a solution in terms of the wavevector ($kd_{\rm i}$) at every frequency ($\omega/\omega_{\rm ci}$) and during every local time interval, and $\omega_{{\rm ci}}$ is the ion cyclotron frequency. Furthermore, we construct the PDF($kd_{\rm i}$, $\omega/\omega_{\rm ci}$) statistically based on the information of $kd_{\rm i}$($\omega/\omega_{\rm ci}$, $t$). The PDFs($kd_{\rm i}$, $\omega/\omega_{\rm ci}$) for the benchmark tests of the Alfv\'en mode and slow mode are illustrated in Figure \ref{fig:fig4}(a) and \ref{fig:fig4}(b). The dispersion relations as indicated by the ridges with high PDF values are fully consistent with the theoretical dispersion relations, which means that the SVD method is well able to resolve the MHD dispersion relations from our observations. Thereafter, we apply the SVD method to the observational measurements to examine whether the dispersion relations of Alfv\'en waves and slow waves prevail. The results are shown in Figure \ref{fig:fig4}(c) and Figure \ref{fig:fig4}(d). The green patches corresponding to high levels of PDF concentrate on and around the theoretical dispersion relations of MHD Alfv\'en and slow modes. These results further confirm the existence of incompressible Alfv\'en waves (the most prevalent component) and compressible slow waves (the sub-dominat component).

\begin{figure}
\plotone{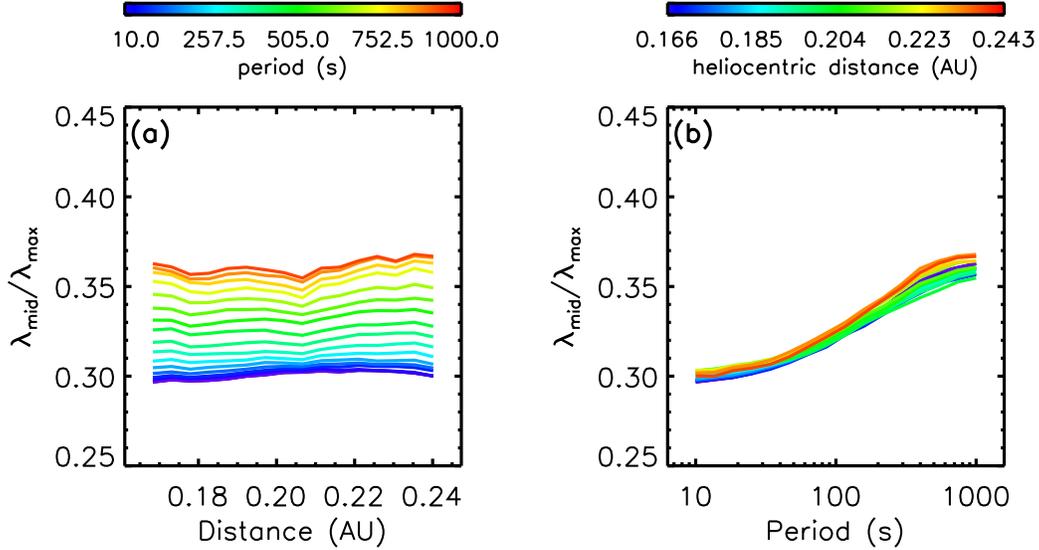}
\caption{(a) The distance profiles of the ratio between the middle and maximum singular values of the magnetic spectral matrix ($\lambda_{{\rm mid}}/\lambda_{{\rm max}}$) for the magnetic fluctuations at different periods from 10 to 1000 s. (b) The variations of $\lambda_{{\rm mid}}/\lambda_{{\rm max}}$ with period for the magnetic fluctuations at different distances from 0.17 to 0.24 AU.
\label{fig:fig5}}
\end{figure}

Lastly, we investigate the variation of the fluctuation anisotropy with period and distance in Figure \ref{fig:fig5}. The ratio of the middle and maximum singular values of the spectral matrix (Eq.(8) of \citet{Santolik2003}), $\lambda_{{\rm mid}}/\lambda_{{\rm max}}$, is adopted to represent the anisotropy of the magnetic field fluctuations in the plane perpendicular to the propagation direction, which we take to be oriented along the direction with the minimum singular value $\lambda_{{\rm min}}$. $\lambda_{{\rm mid}}/\lambda_{{\rm max}}$ is also known as the ellipticity. The ratio increases from around 0.3 to over 0.37 as the wave period increases, throughout the distance range under investigation. According to the above analysis, the dominant Alfv\'en mode increases in its degree of circular or arc polarization with increasing period.

\section{Conclusion}\label{sec:sum}
The diversity, complexity and evolution of solar wind turbulence have always been important research topics in heliospheric physics. Hence, we statistically study mode propagation, mode composition, and fluctuation anisotropy of the solar wind MHD turbulence as measured by PSP. We find that:

(1) At 0.166AU$<$R$<$0.243AU: The propagation angles ($\theta_{{\rm k,B_0}}$) of wave-like turbulent fluctuations for $kd_{\rm i}<0.02$ are greater than $135^{\circ}$, mainly concentrating around $160^{\circ}$, while the distribution gradually shifts its center to $\theta_{{\rm k,B_0}}\sim90^{\circ}$ for $0.02<kd_{\rm i}<0.1$.

(2) The distance variations of the scale-averaged fractions of the MHD modes show that: (a) the outward/anti-sunward propagating Alfv\'en mode dominates the mode composition throughout the whole investigated  range of distances, while the outward slow mode, the inward/sunward Alfv\'en mode and the outward fast mode represent the three smallest proportions; (b) the fraction of the inward fast mode decreases with distance, whereas the fraction of the inward slow mode increases with distance; (c) at 0.166AU$<$R$<$0.215AU, the inward fast mode takes the second place and the inward slow mode is in third place; at 0.215AU$<$R$<$0.243AU, the inward slow mode is in third place followed by the inward fast mode.

(3) The ellipticity increases with spacecraft-frame period in the heliocentric distance range studied.

According to the critical balance hypothesis \citep{Goldreich1995,Horbury2008}, the anisotropy of the power spectrum follows from the condition that the nonlinear time and the propagation time are approximately equal ($k_{\parallel}V_A\sim k_{\perp}\delta v$) in strong MHD turbulence with balanced Els\"asser fluxes. However, in solar wind turbulence with imbalanced fluxes dominated by outward Alfv\'en waves, our probability distribution function of wave propagation in $k_{\parallel}-k_{\perp}$ space (see color maps in Figure \ref{fig:fig2})) is inconsistent with this prediction of critical balance theory (see white solid lines in Figure \ref{fig:fig2}). For $kd_{\rm i}<0.02$, the most probable wavevector is more parallel, while for $kd_{\rm i}>0.02$, the most probable wavevector is closer to the quasi-perpendicular direction. This observational result will help to enlighten and promote the theory of turbulence anisotropy characterized by a transition of propagation direction from quasi-parallel to quasi-perpendicular with a large angular jump at a certain scale. After integrating the ideas of both "slab+2D" and "critical balance" scenarios, an upgraded turbulence phenomenology in Fourier space was proposed to involve "quasi-parallel wavelike fluctuations" and "quasi-2D fluctuations" as well as energy transfer between them and within themselves \citep{Oughton2015}. The observed transition from quasi-parallel to quasi-perpendicular propagation with increasing wavenumber shows a way how to improve the turbulence model in the future.

The transition of the dominant outward Alfv\'en mode from $\theta_{{\rm k,B_0}}\sim160^{\circ}$ to $\theta_{{\rm k,B_0}}\sim90^{\circ}$, as the period decreases from 1000s to 10s, may also indicates the geometry of the Alfv\'en waves at kinetic scales in the near-sun solar wind. Quasi-perpendicular Alfv\'en waves are more likely to dominate at scales closer to ion scale. Accordingly, quasi-perpendicular modes (e.g. kinetic Alfv\'en waves) may participate in the turbulent cascade and further dissipation, energizing and shaping the non-thermal ion distributions, which may develop temperature anisotropic and feed back to excite the ion-cyclotron waves reported during this interval \citep{Bowen2020}. In the future, we will study such chain of energy conversion process: damping of quasi-perpendicular kinetic waves $\longrightarrow$ energization of particles $\longrightarrow$ growth of quasi-parallel waves.

In some respect, our mode composition diagnosis results differ from the results of \citet{Chaston2020}. They study the spectral energy density fractions of six MHD modes inside and outside the field reversal regions, separately. They report that the three outward (anti-sunward) modes are dominant at MHD scales on average. This difference may lie in the calculation of the propagation angle, which is an input parameter of the mode-recognition method \citep{Glassmeier1995}. \citet{Chaston2020} obtained the propagation direction using the spectral matrix of the magnetic field only as suggested by \citet{Samson1980}, while we use both the magnetic and the electric field based on the Faraday's law \citep{Santolik2003}. This aspect may lead to the different results of mode composition.

The ellipticity serves here as an indicator to distinguish if the polarization is circular ($\lambda_{{\rm mid}}/\lambda_{{\rm max}}\sim1$), arc ($0<\lambda_{{\rm mid}}/\lambda_{{\rm max}}<1$), or linear ($\lambda_{{\rm mid}}/\lambda_{{\rm max}}\sim0$). The ellipticity increases with the period of the fluctuations from 0.3 to 0.37, which indicates that the magnetic fluctuations tend to be more and more circular-polarized as the period increases. Based on Figure \ref{fig:fig2}, the waves are also quasi-parallel propagating at larger periods. This observation is consistent with the prediction that the Alfv\'en branch of the MHD solutions is circular-polarized when $\theta_{{\rm k,B_0}}\sim0^{\circ}$.

\section{Acknowledgements}
This work at Peking University (PKU) is supported by NSFC under contracts 41574168, 41674171, 41874200, and 41421003. The team from PKU is also supported by CNSA under contract Nos. D020301 and D020302. D.V. is supported by the STFC Ernest
Rutherford Fellowship ST/P003826/1 and STFC Consolidated Grant ST/S000240/1. S.D.B. acknowledges the
support of the Leverhulme Trust Visiting Professorship program. The authors acknowledge the contributions of the Parker Solar Probe mission operations and spacecraft engineering teams at the Johns Hopkins University Applied Physics Laboratory as well as the FIELDS and SWEAP teams for use of the data. PSP data is available on SPDF (https://cdaweb.sci.gsfc.nasa.gov/index.html/).


\bibliography{PSPMHDwave}
\bibliographystyle{aasjournal}



\end{document}